\documentclass[aps,twocolumn,nofootinbib,showpacs,prd,aps,10pt]{revtex4}
\usepackage[dvips]{graphicx}
\usepackage[english]{babel}
\usepackage[T1]{fontenc}
\usepackage{mathrsfs}
\usepackage[tbtags]{amsmath}
\usepackage{amssymb}
\usepackage{amsxtra}
\usepackage{amsopn}
\usepackage{latexsym}
\usepackage[mathcal]{eucal}

\newcommand{\BE}{\begin{equation}}
\newcommand{\EE}{\end{equation}}
\newcommand{\BA}{\begin{align}}
\newcommand{\EA}{\end{align}}
\newcommand{\Tr}{\mathrm Tr}
\newcommand{\nn}{\nonumber}

\newcommand{\kkdi}{ \frac{i{\rm d}^dk}{(2\pi)^d}}

\newcommand{\qqdi}{ \frac{i{\rm d}^dq}{(2\pi)^d}}
\newcommand{\kkkE}{ \frac{{\rm d}^4k_E}{(2\pi)^4}}
\newcommand{\qqqE}{ \frac{{\rm d}^4q_E}{(2\pi)^4}}
\newcommand{\kkdE}{ \frac{{\rm d}^dk_E}{(2\pi)^d}}

\newcommand{\kkd}{ \frac{{\rm d}^dk}{(2\pi)^d}}
\newcommand{\qqd}{ \frac{{\rm d}^dq}{(2\pi)^d}}
\newcommand{\ttd}{ \frac{{\rm d}^dt}{(2\pi)^d}}

\begin{document}

\title{Second order gluon polarization for SU(N) theory in linear covariant gauge}

\author{Fabio Siringo}

\affiliation{Dipartimento di Fisica e Astronomia 
dell'Universit\`a di Catania,\\ 
INFN Sezione di Catania,
Via S.Sofia 64, I-95123 Catania, Italy}

\date{\today}
\begin{abstract}
The gluon polarization functional is evaluated for a generic linear covariant gauge and for any space dimension
in pure Yang-Mills SU(N) theory up to second order in a generalized perturbation theory, where
the zeroth order action is freely chosen and can be determined by some variational method.
Some numerical data are given for the gluon propagator in Landau gauge and compared with
Feynmann gauge. A comparison is given for several variational methods that can be set up by
the knowledge of the second-order polarization.
\end{abstract}



\maketitle

\section{introduction}

In spite of its phenomenological relevance, the infrared (IR) limit of QCD
has not been fully studied yet because of non-perturbative effects that
limit the power of standard tools based on perturbation theory.
Our current knowledge of the IR limit relies heavily on lattice simulations
while, usually, analitycal non-perturbative techniques can only describe the phenomenology
by insertion of some free parameters that emerge by some unknown sector of the theory
like vertex functions\cite{aguilar8,aguilar9,aguilar10,aguilar13,aguilar14,aguilar14b}, 
counterterms\cite{reinhardt04,reinhardt05,reinhardt08,reinhardt11,reinhardt14,huber14,huber15g,huber15L,huber15b,
huber15c} or renormalization schemes\cite{huber15L}.
A mass parameter for the gluon has been shown to capture most of the non-perturbative effects,
leading to a reasonable fit of lattice data\cite{tissier10,tissier11,tissier14}, 
but we still miss a fully consistent and
analytical {\it ab-initio} theory without spurious fit parameters.

Quite recently, an optimized perturbation theory has been discussed\cite{gep2}, with 
zeroth order trial propagators that are optimized by some variational ansatz.
Many variational strategies can be set up by the simple knowledge of self-energy and
polarization functions, going from the Gaussian effective 
potential\cite{schiff,rosen,barnes,stevenson,stancu2,stancu,ibanez,var,light,superc1,LR,superc2,su2,bubble,HT,AF} 
up to 
Stevenson's minimal sensitivity\cite{minimal} and the novel method of stationary variance\cite{sigma,sigma2,varqed}
that has been shown to be viable for pure Yang-Mills SU(3) in Feynman gauge\cite{varqcd}.
At variance with other analytical approaches, these variational methods have the merit of 
reproducing some lattice features, like the existence of a dynamical mass for the gluon\cite{varqcd},
without any free parameter, because the trial quantities are all optimized by the variational ansatz.
Of course, the agreement with lattice data is not as good as for a fit, but the approximation can
be improved order by order and gives an {\it ab-initio} description that is only based on the original
Lagrangian, without spurious parameters nor undesired counterterms that would spoil the
symmetry of the Lagrangian.  

As discussed in Ref.\cite{gep2}, several variational approaches can be implemented if the
self-energy (the polarization) is known, order by order in the optimized perturbation expansion,
as a functional of the trial propagators.
On the other hand, some internal symmetries of the theory could result broken by the truncated expansion and
would turn out to be only approximately satisfied, so that the actual result would depend on further parameters that have
to do with the gauge choice, the renormalization scheme and even the Renormalization Group (RG) invariance.
All these parameters can also be optimized by a variational ansatz, yielding an optimal gauge or an optimal
renormalization scheme\cite{stevenson2012,stevenson2013}. Thus it would be desirable to have a general 
set of explicit expressions for the polarization functionals,  holding for any gauge, for any renormalization scheme
and for any trial propagator. Actually, most of these functionals have been reported for a free-particle propagator
and in dimensional regularization where many terms vanish. A further proliferation of terms arises from
the use of a generic covariant gauge since the trial propagator would be described by two independent functions
for the transversal and longitudinal part. Despite of many technical problems, the study of a generic linear covariant
gauge has attracted some new interest in the last years and the features of the gluon propagator have been
investigated on the lattice\cite{binosi15,cuccheri10} and in the framework of 
Dyson-Schwinger equations\cite{huber15g,papavassiliou15}. Moreover, it has been recently shown that even if 
some IR properties of the gluon propagator, like the dynamical mass, have no effects in the ultraviolet (UV) 
perturbative regime,
they can drive a quark-quark interaction that is equal to that extracted 
by the ground-state observables\cite{papavassiliou15b}, thus enforcing our interest on the gauge dependence of the gluon
propagator in the IR. 

In this paper we report general integral expressions for the ghost self-energy and the gluon polarization, up
to second order in the optimized perturbation theory, as functionals of trial propagators in a generic linear
covariant gauge, for pure Yang-Mills $SU(N)$ theory in any space dimension $d$. The integral expressions hold
for any renormalization scheme and have been checked by a comparison with known results in dimensional regularization
and in special gauges like Feynman and Landau gauge.
Then, we use that result for extending to Landau gauge a previous calculation of the gluon propagator by the method of
stationary variance\cite{sigma,sigma2,gep2}. 
In fact, the gluon propagator was studied in Feynman gauge in Ref.\cite{varqcd},
while fixed-gauge lattice data are only available in Landau gauge. Here, the numerical results of the calculation are
compared with lattice data in the same gauge and with the outcome of the same method in Feynman gauge.
It turns out that, after renormalization, the gluon propagator is not very sensitive to the gauge change and the
method seems to be approximately gauge invariant, which is a desirable feature of the approximation.
That seems to be in qualitative agreement with Ref.\cite{huber15g}. 
Some different variational methods are discussed and compared, using the same integral expressions 
for the polarization functionals, but
the method of stationary variance emerges as the most reliable among them.

The paper is organized as follows:
in Section II the generalized perturbation theory is reviewed and described in detail for 
the special case of  pure $SU(N)$ Yang-Mills theory; the first-order 
graphs for the polarization are evaluated in Section III; the one-loop second-order graphs are
reported in Section IV, while the two-loop second order graphs are evaluated in Section V (the
expansion is not loop-wise as it is an expansion in powers of the actual interaction);
In Section VI the gluon propagator is evaluated in Landau gauge by the method of stationary variance
and compared with lattice data and with previous results in Feynman gauge;
a discussion and comparison of several variational methods follow in Section VII; some details
on the numerical integration are reported in the appendix.

\section{Generalized perturbation theory}

Let us consider pure Yang-Mills  $SU(N)$ gauge theory without
external fermions. The Lagrangian is
\BE
{\cal L}={\cal L}_{YM}+{\cal L}_{fix}
\EE
where ${\cal L}_{YM}$ is the Yang-Mills term
\BE
{\cal L}_{YM}=-\frac{1}{2} \Tr\left(  \hat F_{\mu\nu}\hat F^{\mu\nu}\right)
\EE
and ${\cal L}_{fix}$ is a guage fixing term.
In terms of the gauge fields, the tensor operator $\hat F_{\mu\nu}$ is given by
\BE
\hat F_{\mu\nu}=\partial_\mu \hat A_\nu-\partial_\nu \hat A_\mu
-i g \left[\hat A_\mu, \hat A_\nu\right]
\EE
where
\BE
\hat A_\mu=\sum_{a} \hat X^a A^a_\mu
\EE
and the generators of $SU(N)$ satisfy the algebra
\BE
\left[ \hat X^a, \hat X^b\right]= i f_{abc} \hat X^c
\EE
with the structure constants normalized according to
\BE
f_{abc} f_{dbc}= N\delta_{ad}.
\label{ff}
\EE
A general covariant gauge-fixing term can be written as
\BE
{\cal L}_{fix}=-\frac{1}{\xi} \Tr\left[(\partial_\mu \hat A^\mu)(\partial_\nu \hat A^\nu)\right]
\EE
and the quantum effective action $\Gamma[A^\prime]$, as a function of an external background field
$A^\prime$ reads 
\BE
e^{i\Gamma[A^\prime]}=\int_{1PI} {\cal D}_{A} e^{iS[A^\prime+A]} J_{FP}[A^\prime+A]
\label{path}
\EE
where $S[A]$ is the action, $J_{FP}[A]$ is the Faddev-Popov determinant
and the path integral represents a sum over one particle irreducible (1PI) graphs\cite{weinbergII}.
Since the gauge symmetry is not broken and we are mainly interested in the propagators,
we will limit to the physical vacuum at $A^\prime=0$,
while a more general formalism can be developed for a full study of the vertex functions\cite{bubble}.

The determinant $J_{FP}$ can be expressed as a path integral over ghost fields
\BE
J_{FP}[A]=\int {\cal D}_{\omega,\omega^\star} e^{iS_{gh}[A,\omega,\omega^\star]} 
\label{pathghost}
\EE
and the effective action can be written as
\BE
e^{i\Gamma}=\int_{1PI} {\cal D}_{A, \omega, \omega^\star} e^{iS_0[A, \omega, \omega^\star]}
e^{iS_I[A, \omega, \omega^\star]}
\label{pathI}
\EE
where the total action in a generic $d$-dimensional space is  
\BE
S_{tot}=\int  {\cal L}_{YM}{\rm d}^dx+\int {\cal L}_{fix}{\rm d}^dx + S_{gh}
\EE
In a generalized perturbation theory\cite{gep2,varqcd}
we have the freedom to split the action in two parts, a trial {\it free} action $S_0$ and the
remaining interaction $S_I$.
We {\it define} the free action $S_0$ as
\begin{align}
S_0&=\frac{1}{2}\int A_{a\mu}(x) {D^{-1}}_{ab}^{\mu\nu}(x,y) A_{b\nu}(y) {\rm d}^dx{\rm d}^dy \nn \\
&+\int \omega^\star_a(x) {G^{-1}}_{ab}(x,y) \omega_b (y) {\rm d}^dx{\rm d}^dy
\label{S0}
\end{align}
where $D^{ab}_{\mu\nu}(x,y)$ and $G_{ab}(x,y)$ are unknown trial matrix functions.
The interaction is then given by the difference 
\BE
S_I=S_{tot}-S_0 
\EE
and can be formally written as the sum of a two-point term and three local terms: the ghost vertex,
the three-gluon vertex and the four-gluon vertex respectively
\BE
S_I=S_2+\int{\rm d}^dx \left[ {\cal L}_{gh} + {\cal L}_3 +   {\cal L}_4\right].
\label{SI}
\EE

In detail, the two-point interaction can be written as
\begin{widetext}
\BE
S_2=\frac{1}{2}\int A_{a\mu}(x) 
\left[{{D_0}^{-1}}_{ab}^{\mu\nu}(x,y)- {D^{-1}}_{ab}^{\mu\nu}(x,y)\right]
A_{b\nu}(y) {\rm d}^dx{\rm d}^dy 
+\int \omega^\star_a(x)
\left[{{G_0}^{-1}}_{ab}(x,y)- {G^{-1}}_{ab}(x,y)\right]
\omega_b (y) {\rm d}^dx{\rm d}^dy
\label{S2}
\EE
\end{widetext}
where $D_0$ and $G_0$ are the standard free-particle propagators for
gluons and ghosts and their Fourier transforms read
\begin{align}
{D_0}_{ab}^{\mu\nu} (p)&=-\frac{\delta_{ab}}{ p^2}\left[t^{\mu\nu}(p)
+\xi \ell^{\mu\nu}(p) \right]\nn\\
{G_0}_{ab} (p)&=\frac{\delta_{ab}}{ p^2}.
\label{D0}
\end{align}
Here the transverse projector $t_{\mu\nu}(p)$ and the longitudinal
projector $\ell_{\mu\nu} (p)$ are defined as
\begin{align}
t_{\mu\nu} (p)&=\eta_{\mu\nu}  - \frac{p_\mu p_\nu}{p^2}\nn\\
\ell_{\mu\nu} (p)&=\frac{p_\mu p_\nu}{p^2}
\label{tl}
\end{align}
and $\eta_{\mu\nu}$ is the metric tensor. 
The three local interaction terms are
\begin{align}
{\cal L}_3&=-g  f_{abc} (\partial_\mu A_{a\nu}) A_b^\mu A_c^\nu\nn\\
{\cal L}_4&=-\frac{1}{4}g^2 f_{abc} f_{ade} A_{b\mu} A_{c\nu} A_d^\mu A_e^\nu\nn\\
{\cal L}_{gh}&=-g f_{abc} (\partial_\mu \omega^\star_a)\omega_b A_c^\mu.
\label{Lint}
\end{align}
The trial functions $G_{ab}$, $D_{ab}^{\mu\nu}$ cancel in the total action $S_{tot}$ which
is exact and cannot depend on them. Thus this formal decomposition holds for any arbitrary
choice of the trial functions and the expansion in
powers of the interaction $S_I$ provides a generalized perturbation theory\cite{gep2,varqcd,varqed}.
Standard Feynman graphs can be drawn for this theory with the trial propagators $D_{ab}^{\mu\nu}$ and 
$G_{ab}$ as free propagators and the vertices that can be read from the
interaction $S_I$ in Eq.(\ref{SI}). As shown in Fig.1, we have two-particle vertices for gluons
and ghosts that arise from the action term $S_2$ in Eq.(\ref{S2}), while the local terms in Eq.(\ref{Lint})
give rise to three- and four-particle vertices.
The effective action $\Gamma$ can be evaluated by perturbation theory as a sum of
Feynman graphs and several variational ansatz can be set up for the best choice of the trial functions\cite{gep2},
mainly relying on stationary conditions that can be easily written in terms of self-energy graphs.
Moreover, the propagators can be written in terms of proper self-energy and polarization functions 
and their evaluation, up to second order, is the main aim of the present paper.
First and second order two-point graphs are shown in  Fig.2.
  
\begin{figure}[t] \label{fig:vertex}
\centering
\includegraphics[width=0.4\textwidth,angle=0]{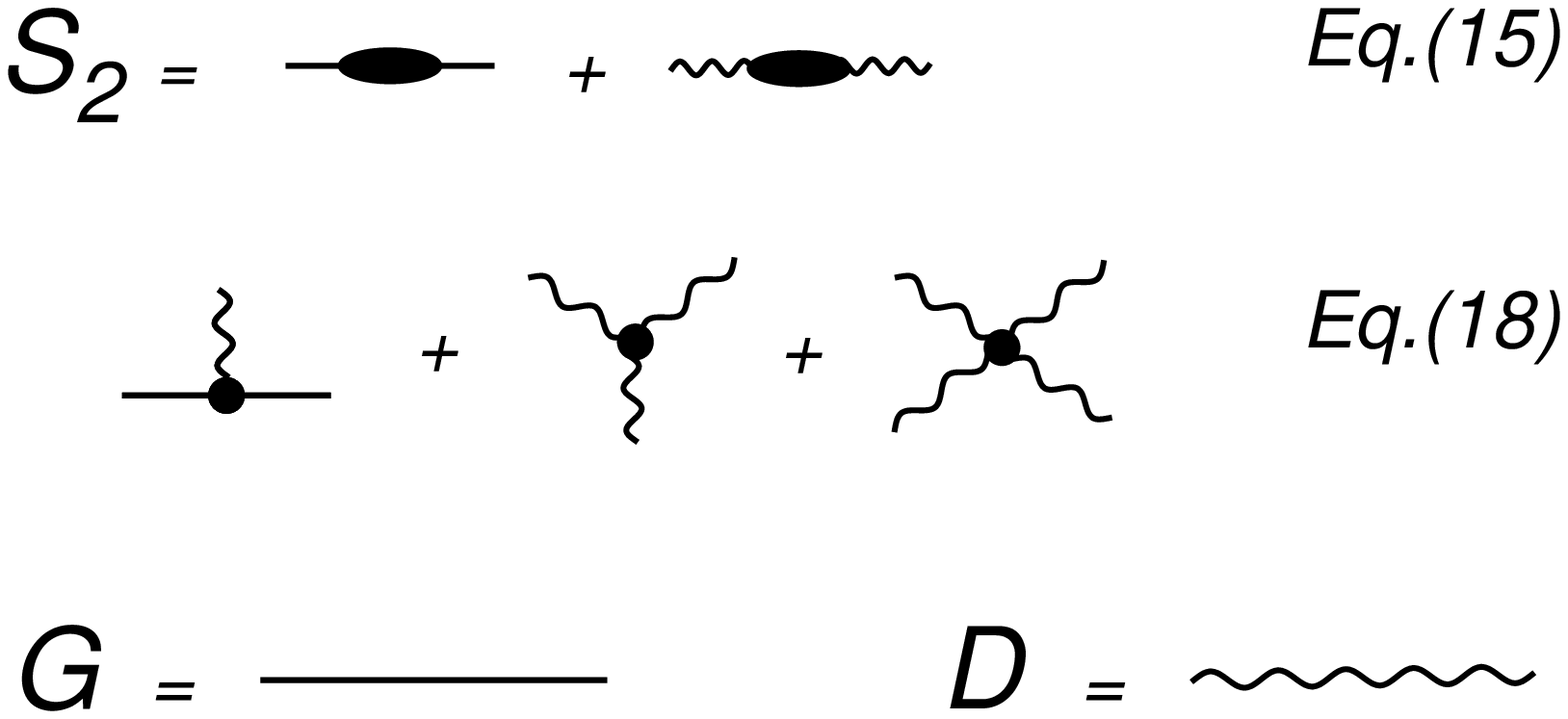}
\caption{The two-point vertices in the interaction $S_2$ of Eq.(\ref{S2}) are shown 
in the first line. The ghost vertex and the three- and four-gluon vertices of Eqs.(\ref{Lint})
are shown in the second line. In the last line the ghost (straight line) and gluon (wavy line)
trial propagators are displayed.}
\end{figure}

\begin{figure}[t] \label{fig:sigma}
\centering
\includegraphics[width=0.29\textwidth,angle=-90]{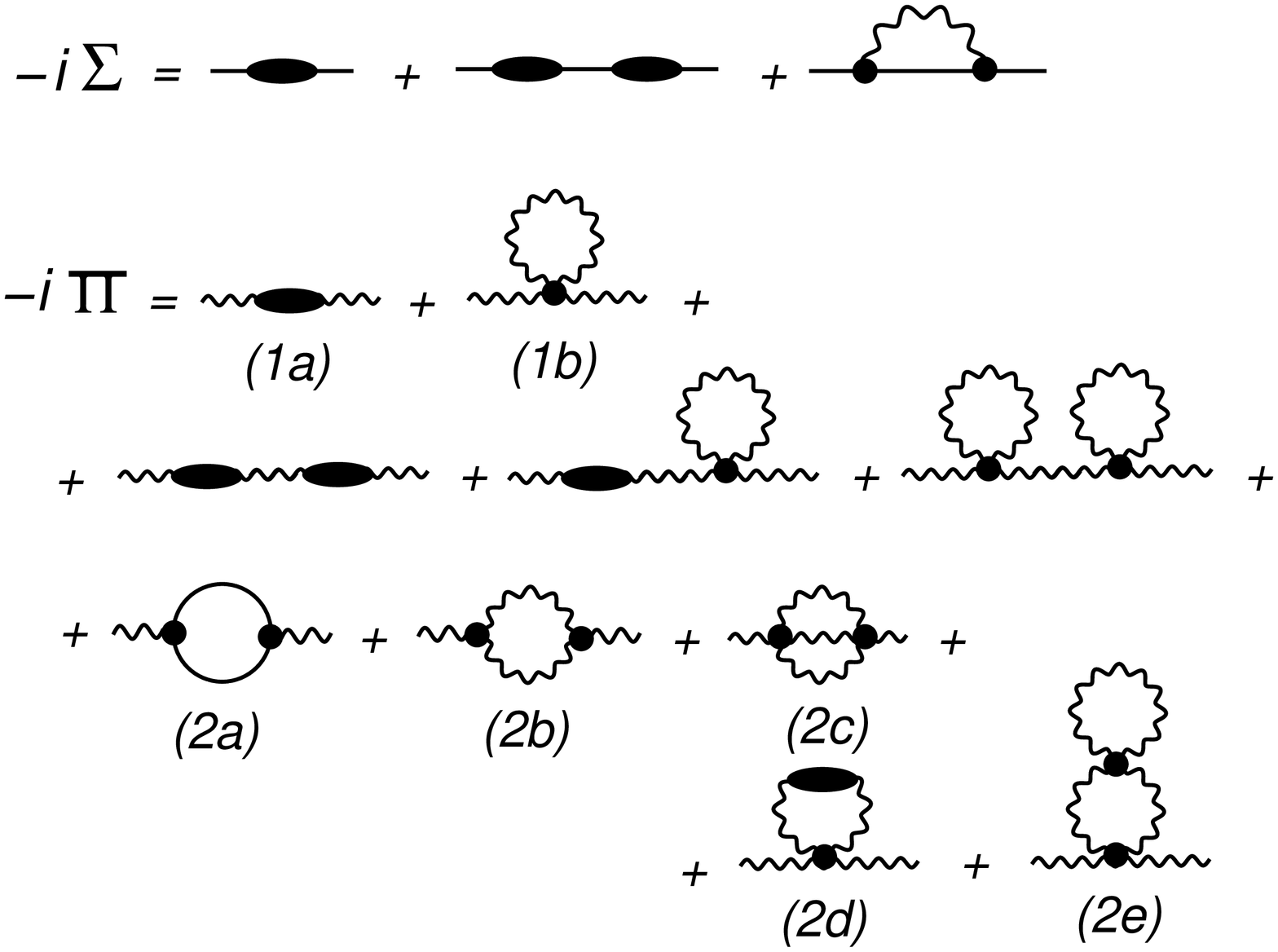}
\caption{First and second order two-point graphs contributing to the ghost self energy and the gluon
polarization. Second order terms include non-irreducible graphs.}
\end{figure}

Since the propagators are gauge-dependent, we write the trial function $D^{ab}_{\mu\nu}$ as the most general
structure that is allowed by Lorentz invariance, namely
\BE
D_{ab}^{\mu\nu} (p)=\delta_{ab}\left[ T(p) t^{\mu\nu}(p)+ L(p) \ell^{\mu\nu}(p)\right]
\label{D}
\EE
while color symmetry ensures that we can always take
\BE
G_{ab} (p)=\delta_{ab} G(p)=\delta_{ab}\frac{\chi (p)}{p^2}
\label{ghdress}
\EE
where $\chi(p)$ is a trial ghost dressing function.
By the same notation, the free-particle propagators in Eq.(\ref{D0}) follow by
inserting in Eq.(\ref{D}) the functions
\BE
T_0(p)=-\frac{1}{p^2}, \>\>L_0(p)=-\frac{\xi}{p^2},\>\> G_0(p)=\frac{1}{p^2}.
\label{free}
\EE
Because of the orthogonality properties of the projectors, the inverse propagator can be
trivially written as 
\BE
{D^{-1}}_{ab}^{\mu\nu} (p)=\delta_{ab}\left[ T(p)^{-1} t^{\mu\nu}(p)+ L(p)^{-1} \ell^{\mu\nu}(p)\right].
\label{D-1}
\EE
The trial propagator of Ref.\cite{varqcd} is recovered in Feynmann gauge ($\xi=1$)
by taking $T(p)=L(p)$, while in Landau gauge ($\xi\to 0$) the longitudinal function $L(p)$ vanishes
and the propagator is transverse.
In both cases the propagator is described by a single function but in the general case two different functions are
required.

\section{First order}

Up to first order, the polarization is given by the sum of graphs $(1a)$ and $(1b)$ in Fig.2.
The tree-graph $\Pi_{(1a)}$ is just
\BE
-i {\Pi_{(1a)}}^{\mu\nu}_{ab}=i{{D_0}^{-1}}_{ab}^{\mu\nu}(p)-i{D^{-1}}_{ab}^{\mu\nu}(p)  
\EE
and in terms of projectors
\BE
{\Pi_{(1a)}}^{\mu\nu}_{ab} (p)=\delta_{ab}[\Pi^{T}_{(1a)}(p)t^{\mu\nu}(p)
+\Pi^{L}_{(1a)}(p) \ell^{\mu\nu}(p)] 
\label{Pproj}
\EE
where
\begin{align}
\Pi^{T}_{(1a)}(p)&=T^{-1}(p)+p^2\nn\\
\Pi^{L}_{(1a)}(p)&=L^{-1}(p)+\frac{p^2}{\xi}.
\end{align}
The one-loop term $\Pi_{(1b)}$ follows from the four-point interaction
term ${\cal L}_4$ in Eq.(\ref{Lint}) that gives the bare vertex
\BE
\Gamma_{abcd}^{\mu\nu\rho\sigma}=-i\frac{g^2}{4!}\left[T_{abcd}^{\mu\nu\rho\sigma}+
T_{acdb}^{\mu\rho\sigma\nu}+T_{adbc}^{\mu\sigma\nu\rho}\right]
\label{Gamma4}
\EE
where the matrix structure $T_{abcd}^{\mu\nu\rho\sigma}$ is a product
of color and Lorentz matrices
\BE
T_{abcd}^{\mu\nu\rho\sigma}=R_{abcd}{S}^{\mu\nu\rho\sigma}
\label{T}
\EE
with 
\BE
R_{abcd}=f_{eab}f_{ecd}
\label{R}
\EE
\BE
S^{\mu\nu\rho\sigma}=\eta^{\mu\rho}\eta^{\nu\sigma}-\eta^{\mu\sigma}\eta^{\nu\rho}.
\label{S}
\EE
The one-loop graph $(1b)$ then reads
\begin{widetext}
\BE
-i {\Pi_{(1b)}}^{\rho\sigma}_{cd}=\frac{4!}{2} \Gamma_{abcd}^{\mu\nu\rho\sigma}
\int\kkd (i\delta_{ab})
\left[T(k)t_{\mu\nu}(k)+L(k)\ell_{\mu\nu}(k)\right]
\EE
and making use of Eq.(\ref{ff}) we can write
\BE
{\Pi_{(1b)}}^{\mu\nu}_{ab}=\delta_{ab} N g^2
\left\{ (d-1)\eta^{\mu\nu}\int\kkdi T(k)
+\int\kkdi\left[L(k)-T(k)\right]t^{\mu\nu}(k)\right\}.
\EE
Integrating in a $d$-dimensional Euclidean space, for a generic function $f(k)$ that only depends on $k^2$,
we can use the identity 
\BE
\int\kkdi \ell_{\mu\nu}(k) f(k)=-\frac{\eta_{\mu\nu}}{d} \int \kkdE f(k_E) \qquad {\rm where}\qquad 
f(k_E)= f(k)\vert_{k^2=-k_E^2}
\EE
\end{widetext}
and write the polarization in terms of the constant integrals 
\begin{align}
I^T_{n,m}&=\int\kkdE [T(k_E)]^n(k_E^2)^m \nn\\
I^L_{n,m}&=\int\kkdE [L(k_E)]^n(k_E^2)^m. 
\label{Inm}
\end{align}
We assume that these diverging integrals are made finite by a regulating scheme to be discussed below.
The one-loop polarization then reads
\BE
{\Pi_{(1b)}}^{\mu\nu}_{ab}=-\delta_{ab}\eta^{\mu\nu} M^2
\label{Pi1b}
\EE
where the first-order mass term $M^2$ is defined as
\BE
M^2=\frac{N g^2 (d-1)}{d}\left[ I^L_{1,0}+(d-1) I^T_{1,0}\right].
\label{gap1}
\EE

It is useful to introduce the transverse and longitudinal
massive functions $T_M(p)$, $L_M(p)$
\begin{align}
[T_M(p)]^{-1}&=[T_0(p)]^{-1}+M^2=-{p^2}+M^2\nn\\
[L_M(p)]^{-1}&=[L_0(p)]^{-1}+M^2=-\frac{p^2}{\xi}+M^2
\label{first}
\end{align}
and the massive propagator
\BE
{D_M}^{\mu\nu}(p)=T_M(p) t^{\mu\nu}(p)+ L_M(p) \ell^{\mu\nu}(p)
\label{DM}
\EE
that describes a free massive particle in a generic covariant gauge.
In the special cases of Feynman gauge ($\xi=1$) and Landau gauge ($\xi\to 0$)
we recover the massive free-particle propagators $D_M^{\mu\nu} (p)=\eta^{\mu\nu}/(-p^2+M^2)$
and $D_M^{\mu\nu} (p)=t^{\mu\nu}(p)/(-p^2+M^2)$, respectively.
With that notation, the total first-order polarization $\Pi_{1}$ can be written in the very simple shape
\BE
{\Pi_{1}}^{\mu\nu}_{ab}={\Pi_{(1a)}}^{\mu\nu}_{ab} +{\Pi_{(1b)}}^{\mu\nu}_{ab}=
{D^{-1}}\>_{ab}^{\mu\nu}-\delta_{ab}{D_M^{-1}}\>^{\mu\nu}.
\label{Pol1}
\EE

There is just one first-order graph for the ghost self-energy, arising from the two-point 
non-local term in Eq.(\ref{S2}) as shown in Fig.2, so that the first-order self-energy can be written
as

\BE
\Sigma^{ab}_1(p)=\delta_{ab}\left[G^{-1}(p)- G^{-1}_0 (p)\right].
\label{Self1}
\EE

The Gaussian Effective Potential 
(GEP)\cite{schiff,rosen,barnes,stevenson,stancu2,stancu,ibanez,var,light,superc1,LR,superc2,su2,bubble,HT,AF}
can be derived by the requirement 
that the functional derivative of the first-order effective potential with respect to the trial functions $D$ and $G$
is zero, that is equivalent\cite{gep2} to the self-consistency condition of a vanishing first-order
self-energy and polarization, $\Pi_1=0$ and $\Sigma_1=0$. The gap equation that arises has been first investigated 
by Cornwall\cite{cornwall82} in 1982 as a simple way to predict a gluon mass.
In the present formalism, from Eqs.(\ref{Pol1}) and (\ref{Self1}) the stationary conditions for the GEP, that derive 
from the vanishing of first-order self-energy and polarization, give a decoupled ghost with $G=G_0$ and
a free massive gluon with $D=D_M$, where the mass $M$ follows from the gauge-dependent gap equation (\ref{gap1})
that can be formally written, by a change of argument in the second integral, 
\BE
M^2=\frac{N (d-1)^2 g_{\xi}^2}{d} \int \kkdE \frac{1}{k_E^2+M^2}
\label{gap2}
\EE
where the gauge dependence has been absorbed by the effective coupling
\BE
g_\xi^2=g^2\left[ 1+\frac{\xi^{d/2}}{d-1}\right].
\EE
According, in Feynman gauge the larger effective coupling would give a larger mass as compared with Landau gauge.

\section{Second order - one loop}

The generalized perturbation theory that arises from the expansion in powers of the interaction $S_I$ is
not a loopwise expansion in powers of the coupling constant, so that one-loop and two-loop graphs coexist
in the second order term of the polarization. The standard one-loop graphs, namely the ghost and gluon 
one-loop graphs, $\Pi_{(2a)}$ and $\Pi_{(2b)}$ in Fig.2, are described in this section together with 
the one-loop ghost self-energy. The other second-order terms, namely
the second-order one-loop graph $\Pi_{(2d)}$ and the two-loop graphs $\Pi_{(2c)}$ and $\Pi_{(2e)}$ 
will be discussed in the next section.

\subsection{One-loop ghost self-energy}
The one-loop ghost self-energy follows from the bare vertex of
the ghost-gluon interaction term ${\cal L}_{gh}$ in Eq.(\ref{Lint}) 
\begin{widetext}
\BE
-i {\Sigma}_{ab} (p)=g^2 f_{cda} f_{c^\prime bd^\prime }\int\kkd (p_\mu-k_\mu) p_\nu
\left[ i D^{\mu\nu}_{c c^\prime} (k)
i G_{d d^\prime} (p-k)\right].
\EE
Making use of Eq.(\ref{ff}) and integrating in a $d$-dimensional Euclidean space,
we can split the self-energy in two terms
\BE
{\Sigma}_{ab} (p) = \delta_{ab}\left[\Sigma^T (p)+\Sigma^L (p)\right]
\label{SigmaTL}
\EE
where
\begin{align}
\Sigma^{T} (p)&=-Ng^2
\int\kkdE\>\frac{\chi(k_E-p_E) T(k_E)}{(k_E-p_E)^2}
\left[p_E^2 -\frac{(k_E\cdot p_E)^2}{k_E^2}\right]
\nn\\
\Sigma^{L} (p)&=-Ng^2
\int\kkdE\>\frac{\chi(k_E-p_E) L(k_E)}{(k_E-p_E)^2}
\left[ \frac{(k_E\cdot p_E)^2}{k_E^2}-(k_E\cdot p_E)\right].
\label{SigmaTL2}
\end{align}
These integrals are functionals of the ghost dressing function $\chi$ of Eq.(\ref{ghdress})
and of the gauge-dependent transverse and longitudinal gluon propagators, respectively.

\subsection{Ghost loop}
The one-loop polarization term $\Pi_{(2a)}$, the ghost loop in Fig.2, also follows from the bare vertex of
the ghost-gluon interaction term ${\cal L}_{gh}$ in Eq.(\ref{Lint}) 
\BE
-i {\Pi_{(2a)}}^{\mu\nu}_{cd}(p)=-g^2 f_{abc} f_{bad}
\int\kkd (p+k)^\mu k^\nu
i G(p+k) iG(k).
\EE
Making use of Eq.(\ref{ff}) and integrating in a $d$-dimensional Euclidean space,
with the same notation of Eq.(\ref{Pproj}), we can write the transverse and longitudinal
parts in terms of the trial ghost dressing function $\chi$ of Eq.(\ref{ghdress})
\begin{align}
\Pi^{T}_{(2a)}(p)&=-\frac{Ng^2}{(d-1)}
\int\kkdE\frac{\chi(p_E+k_E)\chi(k_E)}{(p_E+k_E)^2}
\left[1-\frac{(k_E\cdot p_E)^2}{k_E^2p_E^2}\right]
\nn\\
\Pi^{L}_{(2a)}(p)&=-Ng^2
\int\kkdE\frac{\chi(p_E+k_E)\chi(k_E)}{(p_E+k_E)^2k_E^2}
\left[(k_E\cdot p_E)+\frac{(k_E\cdot p_E)^2}{p_E^2}\right]
\end{align}
in agreement with the result reported by other authors\cite{reinhardt14}.

\subsection{Gluon loop}
The one-loop polarization term $\Pi_{(2b)}$, the gluon loop in Fig.2, follows from 
the gluon-gluon interaction term ${\cal L}_{3}$ in Eq.(\ref{Lint}) that gives 
the bare three-particle vertex
\BE
\Gamma_{abc}^{\mu\nu\rho}(p,q,k)=
\frac{g f_{abc}}{3!}\left\{\eta^{\mu\nu} (p^\rho-q^\rho)+
\eta^{\rho\nu} (q^\mu-k^\mu)+\eta^{\mu\rho} (k^\nu-p^\nu)\right\}.
\EE
The one-loop graph $(2b)$ then reads
\BE
-i {\Pi_{(2b)}}^{\mu\nu}_{aa^\prime}(p)=\frac{3!3!}{2} 
\int\kkd 
i D_{b b^\prime, \rho\rho^\prime} (p+k) 
iD_{c c^\prime, \tau\tau^\prime} (k)
\Gamma_{abc}^{\mu\rho\tau}(p,-p-k,k) 
\Gamma_{a^\prime b^\prime c^\prime}^{\nu\rho^\prime\tau^\prime}(-p,p+k,-k)
\label{Pi2b}
\EE
\end{widetext}
and since the trial propagator is defined by two independent functions, there
are 36 terms for each of the longitudinal and transverse parts of the polarization.
We can write them in a more compact shape by introducing some degree of redundancy in the
notation. By Eq.(\ref{ff}), the sum over color indices gives a diagonal matrix, so that we
can use the same notation of Eq.(\ref{Pproj}) and drop all color indices.
Let us denote by $\alpha,\beta,\gamma$ the three momenta in the vertex,
$\alpha=-p$, $\beta=p+k$, $\gamma=-k$, so that $\alpha+\beta+\gamma=0$. Then, denote by 
${\hat A_a}^{\mu\nu},{\hat B_b}^{\mu\nu},{\hat C_c}^{\mu\nu}$ the three projectors
\begin{align}
{\hat A_a}^{\mu\nu}&={P_a}^{\mu\nu} (\alpha) \nn\\
{\hat B_b}^{\mu\nu}&={P_b}^{\mu\nu} (\beta) \nn\\
{\hat C_c}^{\mu\nu}&={P_c}^{\mu\nu} (\gamma) 
\label{Proj}
\end{align}
where $a,b,c=\pm 1$, while $P_{\pm}^{\mu\nu}$ are the transverse and longitudinal projectors 
$P_{+}^{\mu\nu} (k)=t^{\mu\nu} (k)$ and
${P_{-}}^{\mu\nu} (k)=\ell^{\mu\nu} (k)$, that
can also be written as
\BE
P_{a}^{\mu\nu} (k)=n_a\eta^{\mu\nu} -a \ell^{\mu\nu} (k)
\EE
where $n_a=(1+a)/2$.

Moreover, with the same notation of Eq.(\ref{Pproj}), let us denote by ${\cal A}_a, {\cal B}_b, {\cal C}_c$ the numbers
\begin{align}
{\cal A}_+&=(d-1) \Pi^T_{(2b)} (\alpha), \qquad {\cal A}_-=\Pi^L_{(2b)} (\alpha) \nn\\
{\cal B}_+&= T(\beta), \qquad {\cal B}_-= L(\beta)\nn\\
{\cal C}_+&= T(\gamma), \qquad {\cal C}_-= L(\gamma)
\end{align}
so that, having dropped color indices (not to be confused with the sign indices $a,b,c$ in this section), we can write
\begin{align}
{\cal A}_a&= {\hat A_a}^{\mu\nu}\> {\Pi_{(2b)}}_{\mu\nu} (p) \nn\\
D^{\mu\nu} (p+k)&=\sum_{b=\pm 1} {\cal B}_b{\hat B_b}^{\mu\nu} \nn\\
D^{\mu\nu} (k)&=\sum_{c=\pm 1} {\cal C}_c{\hat C_c}^{\mu\nu}. 
\end{align}
Inserting in Eq.(\ref{Pi2b}), the transverse and longitudinal polarizations ${\cal A}_a$ can be simply
written as
\BE
{\cal A}_a=\frac{Ng^2}{2} \sum_{bc}\int\kkdi {\cal B}_b {\cal C}_c \>{\cal F}_{abc} (\alpha,\beta,\gamma)
\label{Pi2bA}
\EE
where, with the obvious shorthand notation
\begin{widetext}
\BE
\left[k\cdot \hat X\cdots \hat Y \cdot p\right]
=k_\mu {{\hat X}^\mu}_{\>\>\rho} \cdots {{\hat Y}^\tau}_{\>\>\nu} p^\nu,\qquad  
\left[\hat X\cdot \hat Y\right]={\hat X}^{\mu\nu} {\hat Y}_{\nu\mu},  
\label{short} 
\EE
the matrix ${\cal F}$ reads
\BE
{\cal F}_{abc} (\alpha,\beta,\gamma)=\left[(\alpha-\beta)\cdot{\hat C}_c\cdot (\alpha-\beta)\right]
\left[{\hat A}_a\cdot {\hat B}_b\right]
+2\left[(\alpha-\beta)\cdot{\hat C}_c\cdot {\hat B}_b\cdot {\hat A}_a\cdot(\beta-\gamma)\right]
+ {\rm cycl.}\>{\rm perm.},
\EE
summed over the three cyclic simultaneous permutations of all the arguments, indices and projectors,
i.e.  $\alpha\to\beta\to \gamma\to \alpha$ together with $\>a\to b\to c\to a$ and 
${\hat A}\to {\hat B} \to {\hat C} \to {\hat A}$.
A straightforward but tedious calculation yields
\begin{align}
{\cal F}_{abc} (\alpha,\beta,\gamma)&=3 (n_a n_b n_c) (d-1)(\alpha^2+\beta^2+\gamma^2)
+\left\{(acn_b)\beta^2\left[1-\frac{(\beta^2-\alpha^2-\gamma^2)^2}{4\alpha^2\gamma^2}\right]
+ {\rm cycl.}\>{\rm perm.}\right\}\nn\\
&+\left\{(an_bn_c)\left[\frac{\alpha^2}{2}-(\beta^2+\gamma^2)-(2d-3)\frac{(\beta^2-\gamma^2)^2}{2\alpha^2}\right]
+ {\rm cycl.}\>{\rm perm.}\right\}.
\label{kernel}
\end{align}
\end{widetext}
The result holds for any gauge parmeter $\xi$, space dimension $d$, and for any trial propagator. 
It has been found in agreement with previous calculations in Feynman gauge\cite{varqcd} and  
with older results for a free-particle trial propagator in generic covariant gauge\cite{watson}.

For instance, in Feynman gauge, the polarization function $\Pi (p)$ of Ref.\cite{varqcd} is defined
as $\Pi (p)=({\cal A}_+ +{\cal A}_-)/d$ and the trial propagator is taken to be $T(k)=L(k)$ so that
the kernel of the integral in Eq.(\ref{Pi2bA}) is just given by
\BE
\frac{1}{d} \sum_{abc}{\cal F}_{abc} (\alpha,\beta,\gamma)=\frac{6(d-1)}{d}(p^2+k^2+p\cdot k)
\EE
in agreement with Eq.(A11) of Ref.\cite{varqcd} for $d=4$.
In the same work the function $\Pi^\prime (p)$ is the transverse polarization 
$\Pi^\prime (p)={\cal A}_+/(d-1)$ and the corresponding kernel in Eq.(\ref{Pi2bA}), for $d=4$, is given by
\begin{align}
\frac{1}{d-1} \sum_{bc}\left[{\cal F}_{abc} (\alpha,\beta,\gamma)\right]_{a=1}&=
5p^2+2k^2+2p\cdot k\nn\\
&+\frac{10}{3}k^2\left[1-\frac{(p\cdot k)^2}{k^2 p^2}\right]
\end{align}
in agreement with Eq.(A12) of Ref.\cite{varqcd}.
In the work of Watson\cite{watson} the trial propagator is taken to be the gauge-dependent
free-particle function $D_0$ as defined in Eqs.(\ref{D0}), (\ref{free}). 
The function $\hat J_p^{(1)}$ in Eq.(3.3.4) of
that work\cite{watson} is defined as $(d-1)\hat J_p^{(1)}=({\cal A}_+ +{\cal A}_-)/p^2$ and the
corresponding kernel in Eq.(\ref{Pi2bA}) can be written as a polynomial 
$\left[w_0+w_1\xi+w_2\xi^2\right]$ with coefficients
\begin{align}
w_0&=\frac{1}{\alpha^2}\sum_a \left[{\cal F}_{abc} (\alpha,\beta,\gamma)\right]_{b=c=1}
\nn\\
w_1&=\frac{1}{\alpha^2}\sum_a \left\{
\left[{\cal F}_{abc} (\alpha,\beta,\gamma)\right]_{\begin{smallmatrix} b=+1\\c=-1\end{smallmatrix}}+
\left[{\cal F}_{abc} (\alpha,\beta,\gamma)\right]_{\begin{smallmatrix} b=-1\\c=+1\end{smallmatrix}}\right\}
\nn\\
w_2&=\frac{1}{\alpha^2}\sum_a \left[{\cal F}_{abc} (\alpha,\beta,\gamma)\right]_{b=c=-1}.
\end{align}
The coefficients can be easily evaluated by Eq.(\ref{kernel}) and if we drop all terms that vanish under
integration (because of symmetry or by dimensional regularization) we obtain
\begin{align}
w_0&=3d-\frac{7}{2}+2(2-d)\frac{\alpha^2}{\gamma^2}-\frac{\alpha^4}{4\gamma^2\beta^2}\nn\\
w_1&=1+(2d-5)\frac{\alpha^2}{\gamma^2}+\frac{\alpha^4}{2\gamma^2\beta^2}\nn\\
w_2&=\frac{1}{2}+\frac{\alpha^2}{\gamma^2}-\frac{\alpha^4}{4\gamma^2\beta^2}
\end{align}
in agreement with Eq.(3.3.4) of Ref.\cite{watson} that was evaluated by a computer routine
for algebraic computations. The general result in Eq.(\ref{kernel}) holds for
any choice of the trial propagator and contains all terms that might not vanish by symmetry or
dimensional regularization when a generic massive propagator is considered.
Moreover the result does not depend on a specific regularization scheme and can be used for any
kind of calculation.

In Landau gauge, the propagator is transverse and is defined by only one function $T(p)$, so that
the transverse polarization ${\cal A}_+$ is obtained by retaining only one term, for $b=c=+1$,
in Eq.(\ref{Pi2bA}), and the corresponding kernel for $d=4$ reads
\begin{align}
&\left[ {\cal F}_{abc} (\alpha,\beta,\gamma)\right]_{a=b=c=+1}=\left[1-\frac{(k\cdot p)^2}{k^2p^2}\right]\times\nn\\
&\times\left[11(k^2+p^2)+2(k\cdot p)+\frac{p^4+10p^2 k^2+k^4)}{(k+p)^2}\right].
\end{align}

\section{Second order - two loop}

Besides the standard one-loop graphs of the previous section, the second-order polarization includes
the one-loop and two-loop tadpoles $\Pi_{(2d)}$, $\Pi_{(2e)}$ and the two-loop sunset $\Pi_{(2c)}$.

\subsection{Tadpoles}

The one-loop and two-loop graphs, $\Pi_{(2d)}$ and  $\Pi_{(2e)}$ in Fig.2, follow from the standard tadpole
$\Pi_{(1b)}$ by insertion of the total first-order polarization  in the loop
\BE
iD^{\mu\nu}\to iD^{\mu\rho}(-i {\Pi_1}_{\rho\sigma}) iD^{\sigma\nu}
\EE
that is, making use of Eq.(\ref{Pol1}),
\BE
D^{\mu\nu}\to D^{\mu\nu}-D^{\mu\rho} {D^{-1}_M}_{\rho\sigma} D^{\sigma\nu}.
\EE
Insertion in Eqs.(\ref{Pi1b}),(\ref{gap1}) yields 
\BE
{\Pi_{(2d)}}^{\mu\nu}_{ab}+{\Pi_{(2e)}}^{\mu\nu}_{ab}
=-\delta_{ab}\eta^{\mu\nu} \left[ M^2-{\cal M}^2\right]
\EE
where the new mass constant reads
\BE
{\cal M}^2=\frac{N g^2 (d-1)}{d}\left[ M^2I^L_{2,0}+\frac{I^L_{2,1}}{\xi}
+(d-1)(M^2 I^T_{2,0} +I^T_{2,1})\right].
\EE
The constant integrals $I^T_{n,m}$, $I^L_{n,m}$ were defined
in Eq.(\ref{Inm}) and are functionals of the transverse and longitudinal
trial functions $T(p)$, $L(p)$, respectively.

\subsection{Two-loop sunset}

The two-loop graph $(2c)$ in Fig.2 is the most involved and even if its 
contribution is small when the coupling is not too large, it can be relevant
in a variational approach when the coupling is allowed to increase enough.
The calculation follows from the explicit four-gluon bare vertex 
in Eq.(\ref{Gamma4})
\begin{widetext}
\BE
-i {\Pi_{(2c)}}^{\mu\mu^\prime}_{aa^\prime}(p)=\frac{4!4!}{3!} 
\Gamma_{abce}^{\mu\nu\rho\sigma}
\Gamma_{a^\prime b^\prime c^\prime e^\prime}^{\mu^\prime\nu^\prime\rho^\prime\sigma^\prime}
\int\kkd \int\qqd \int\ttd
\left[i D_{b b^\prime, \nu\nu^\prime} (k) \right]
\left[iD_{c c^\prime, \rho\rho^\prime} (q)\right]
\left[iD_{e e^\prime, \sigma\sigma^\prime} (t)\right](2\pi)^d\delta^d(k+q+t+p)
\label{Pi2c}
\EE
and the explicit symmetry by permutation of dummy integration variables ensures that,
under integration,  the whole expression is symmetric for the exchange of the corresponding Lorentz and
color indices in the matrix factors. Thus the sum over permutation in Eq.(\ref{Gamma4}) can be replaced by
a factor of three in one of the vertices, yielding
\BE
{\Pi_{(2c)}}^{\mu\mu^\prime}_{aa^\prime}(p)=\frac{g^4}{2} 
T_{abce}^{\mu\nu\rho\sigma}
\left[
T_{a^\prime b c e}^{\mu^\prime\nu^\prime\rho^\prime\sigma^\prime}
+T_{a^\prime c e b}^{\mu^\prime\rho^\prime\sigma^\prime\nu^\prime}
+T_{a^\prime e b c}^{\mu^\prime\sigma^\prime\nu^\prime\rho^\prime}
\right]
\sum_{ijk=\pm 1}\int\kkdi \int\qqdi 
\hat B_i^{\sigma\sigma^\prime}
\hat C_j^{\rho\rho^\prime}
\hat E_k^{\nu\nu^\prime} {\cal B}_i{\cal C}_j{\cal E}_k
\label{Pi2c2}
\EE
\end{widetext}
with a compact notation that extends that of the previous section: here we define the four vectors 
$\alpha=p$, $\beta=-(p+q+k)$, $\gamma=q$, $\epsilon=k$ so that $\alpha+\beta+\gamma+\epsilon=0$
and add a new projector to the set in Eq.(\ref{Proj}) 
\BE
{\hat E_e}^{\mu\nu}={P_e}^{\mu\nu} (\epsilon). 
\EE
Moreover, in this section, we denote by ${\cal A}_a, {\cal B}_b, {\cal C}_c, {\cal E}_e$ the numbers
\begin{align}
{\cal A}_+&=(d-1) \Pi^T_{(2c)} (\alpha), \qquad {\cal A}_-=\Pi^L_{(2c)} (\alpha) \nn\\
{\cal B}_+&= T(\beta), \qquad {\cal B}_-= L(\beta)\nn\\
{\cal C}_+&= T(\gamma), \qquad {\cal C}_-= L(\gamma)\nn\\
{\cal E}_+&= T(\epsilon), \qquad {\cal E}_-= L(\epsilon)
\end{align}
so that, dropping color indices, we can write
\begin{align}
{\cal A}_a&= {\hat A_a}^{\mu\nu}\> {\Pi_{(2c)}}_{\mu\nu} (p) \nn\\
D^{\mu\nu} (p+q+k)&=\sum_{b=\pm 1} {\cal B}_b{\hat B_b}^{\mu\nu} \nn\\
D^{\mu\nu} (q)&=\sum_{c=\pm 1} {\cal C}_c{\hat C_c}^{\mu\nu}\nn\\ 
D^{\mu\nu} (k)&=\sum_{e=\pm 1} {\cal E}_c{\hat E_e}^{\mu\nu}. 
\label{functions}
\end{align}
Under integration, the matrix structure in Eq.(\ref{Pi2c2}) simplifies because of the permutation
symmetry of dummy integration variables and the three matrix products can be recast as a single
Lorentz matrix $\omega^{\mu\mu^\prime}$ that multiplies three color matrices
\begin{align}
{\Pi_{(2c)}}^{\mu\mu^\prime}_{aa^\prime}&=
\omega^{\mu\mu^\prime}\left[2R_{a^\prime b c e}R_{a^\prime b c e}
-R_{a^\prime b c e}(R_{a^\prime c e b}+R_{a^\prime e b c})\right]\nn\\
&=3N^2\delta_{a a^\prime}\omega^{\mu\mu^\prime}
\end{align}
where the last equality folows by Jacobi identity.
Then, dropping color indices, the transverse and longitudinal polarizations
${\cal A}_a$ can be written as
\BE
{\cal A}_a=\frac{3N^2g^4}{2} \sum_{bce}\int\kkdi\int\qqdi 
{\cal B}_b {\cal C}_c {\cal E}_e \>{\cal F}_{abce} (\alpha,\beta,\gamma,\epsilon)
\label{Pi2cA}
\EE
where the kernel ${\cal F}$ follows from the projection of $\omega^{\mu \mu^\prime}$
by the projector $\hat A$ according to Eq.(\ref{functions}) and with the shorthand
notation of Eq.(\ref{short}) it can be written in terms of traces of projectors
\BE
{\cal F}_{abce} (\alpha,\beta,\gamma, \epsilon)=\left[{\hat A}_a\cdot{\hat E}_e\right]
\left[{\hat C}_c\cdot{\hat B}_b\right]-
\left[{\hat A}_a\cdot {\hat E}_b \cdot {\hat C}_a\cdot {\hat B}_b\right].
\EE
Because of the obvious symmetry of the integral in Eq.(\ref{Pi2cA}), the result is invariant
for any simultaneous permutations of the last three arguments, indices and projectors,
i.e.  $\beta\to \gamma\to \epsilon \to \beta$ together with $b\to c\to e \to  b$ and 
${\hat B} \to {\hat C} \to {\hat E} \to {\hat B}$. Using that symmetry the kernel
${\cal F}$ can be written as
\begin{widetext}
\begin{align}
&{\cal F}_{abce} (\alpha,\beta,\gamma, \epsilon)=d(d-1) (n_a n_b n_c n_e)-3(d-1)(n_a b n_c n_e)
-(d-1)(a n_b n_c n_e)
+(n_a n_b c e)\left[2+(d-3)\frac{(\epsilon\cdot\gamma)^2}{\epsilon^2 \gamma^2}\right]\nn\\
&+(a n_b c n_e)\left[2+(d-3)\frac{(\alpha\cdot\gamma)^2}{\alpha^2 \gamma^2}\right]
-(n_a b c e)\frac{(\epsilon\cdot\gamma)}{\epsilon^2 \gamma^2}
\left[(\epsilon\cdot\gamma)-\frac{(\beta\cdot\gamma)(\beta\cdot\epsilon)}{\beta^2}\right]
+(an_b c e)\frac{(\epsilon\cdot\gamma)}{\epsilon^2 \gamma^2}
\left[(\epsilon\cdot\gamma)-\frac{(\alpha\cdot\gamma)(\alpha\cdot\epsilon)}{\alpha^2}\right]\nn\\
&+2(ab n_c e)\frac{(\alpha\cdot\beta)}{\alpha^2 \beta^2}
\left[(\alpha\cdot\beta)-\frac{(\alpha\cdot\epsilon)(\beta\cdot\epsilon)}{\epsilon^2}\right]
+(abce)\frac{(\alpha\cdot\beta)(\gamma\cdot\epsilon)}{\alpha^2 \beta^2 \gamma^2 \epsilon^2}
\left[(\alpha\cdot\beta)(\gamma\cdot\epsilon)-(\alpha\cdot\epsilon)(\beta\cdot\gamma)\right].
\label{kernel2}
\end{align}
\end{widetext}
That explicit expression gives the two-loop sunset graph $(2c)$ by a double integration
in Eq.(\ref{Pi2cA}) and holds for any covariant gauge, any trial propagator and any space dimension.
For instance, in Feynman gauge, the polarization function $\Pi (p)$ of Ref.\cite{varqcd} is defined
as $\Pi (p)=({\cal A}_+ +{\cal A}_-)/d$ and the trial propagator is taken to be $T(k)=L(k)$ so that
the kernel of the integral in Eq.(\ref{Pi2cA}) is a constant and is just given by
\BE
\frac{1}{d} \sum_{abce}{\cal F}_{abce} (\alpha,\beta,\gamma,\epsilon)=(d-1)
\EE
in agreement with Eq.(A19) of Ref.\cite{varqcd} for $d=4$.
In the same work the function $\Pi^\prime (p)$ is the transverse polarization 
$\Pi^\prime (p)={\cal A}_+/(d-1)$ and the corresponding kernel in Eq.(\ref{Pi2cA}) is given by
\BE
\frac{1}{d-1} \sum_{bce}\left[{\cal F}_{abce} (\alpha,\beta,\gamma, \epsilon)\right]_{a=1}=(d-1)
\EE
again in agreement with Eq.(A19) of Ref.\cite{varqcd}.

In Landau gauge the result is more involved. The propagator is transverse and is defined by one function $T(p)$, 
so that the transverse polarization ${\cal A}_+$ is obtained by retaining only one term for $b=c=e=+1$
in Eq.(\ref{Pi2cA}). The corresponding kernel for $d=4$ reads
\begin{widetext}
\begin{align}
&\left[ {\cal F}_{abce} (\alpha,\beta,\gamma, \epsilon)\right]_{a=b=c=e=+1}=4+
\frac{(k\cdot q)}{k^2q^2}\left[\frac{\left[q\cdot (k+q+p)\right]\left[k\cdot (k+q+p)\right]}{(k+q+p)^2}
-\frac{(p\cdot q)(p\cdot k)}{p^2}\right]\nn\\
&+\frac{(k\cdot q)^2}{k^2q^2}+\frac{(p\cdot q)^2}{p^2q^2}+\frac{2\left[p\cdot(p+q+k)\right]}{p^2(p+q+k)^2}
\left[p\cdot(p+q+k)-  \frac{(p\cdot k)\left[k\cdot (k+q+p)\right]}{k^2}\right]\nn\\
&+\frac{(k\cdot q)\left[p\cdot(p+q+k)\right]}{p^2q^2 k^2(p+q+k)^2}
\Big[
\left[p\cdot(p+q+k)\right](k\cdot q)-(p\cdot k)\left[q\cdot (k+q+p)\right]
\Big].
\label{Landau2c}
\end{align}
\end{widetext}

\section{Stationary Variance in Landau gauge}

Explicit expressions for the second-order graphs are useful for a direct comparison of variational results
in different gauges. For instance, the method of stationary variance\cite{sigma,sigma2,gep2,varqed} has 
been shown to be viable in Feynman gauge where provides a good description of the gluon propagator\cite{varqcd}.
Here, we explore the outcome of the same method in Landau gauge where lattice simulations are
available\cite{bogolubsky,dudal}. 

The variance is stationary when the trial propagators satisfy the stationary conditions\cite{gep2}
\begin{align}
\Pi_2&=\Pi_1\nn\\
\Sigma_2&=\Sigma_1
\label{variance}
\end{align}
where $\Sigma_1$ is the first-order ghost self-energy in Eq.(\ref{Self1}), 
$\Sigma_2$ is the sum of reducible and irreducible second-order graphs,
$\Pi_1$ is the sum of
the first-order polarization graphs $(1a)$, $(1b)$ and $\Pi_2$ is the sum of
reducible and irreducible second-order polarization graphs.
With the notation of Eqs.(\ref{Pol1}),(\ref{Self1}), (\ref{SigmaTL}) the stationary equations read
\begin{widetext}
\begin{align}
{D^{-1}}\>_{ab}^{\mu\nu}-\delta_{ab}{D_M^{-1}}\>^{\mu\nu}&=
\left[ {D^{-1}}\>_{ac}^{\mu\rho}-\delta_{ac}{D_M^{-1}}\>^{\mu\rho} \right]
\left[D_{ce, \rho\sigma}\right]
\left[ {D^{-1}}\>_{eb}^{\sigma\nu}-\delta_{eb}{D_M^{-1}}\>^{\sigma\nu} \right]
+{\Pi^\star_2}\>_{ab}^{\mu\nu}\nn\\
{G^{-1}}-{G_0^{-1}}&=
\left[{G^{-1}}-{G_0^{-1}}\right]
G
\left[{G^{-1}}-{G_0^{-1}}\right]
+{\Sigma^T}+{\Sigma^L}
\label{Stat}
\end{align}
where the proper polarization $\Pi^\star_2$ is the sum of all the irreducible second-order terms
\BE
{\Pi^\star_2}\>_{ab}^{\mu\nu}=
{\Pi_{(2a)}}\>_{ab}^{\mu\nu}+
{\Pi_{(2b)}}\>_{ab}^{\mu\nu}+
{\Pi_{(2c)}}\>_{ab}^{\mu\nu}+
{\Pi_{(2d)}}\>_{ab}^{\mu\nu}+
{\Pi_{(2e)}}\>_{ab}^{\mu\nu}.
\label{Pistar}
\EE
\end{widetext}
The coupled set of integral equations (\ref{Stat}) can be written as
\begin{align}
D\>_{ab}^{\mu\nu}&=\delta_{ab}{D_M}\>^{\mu\nu}-{D_M}\>^{\mu\rho} {\Pi^\star_2}\>_{ab, \rho\sigma}
{D_M}\>^{\sigma\nu} \nn\\
G&=G_0-G_0\left[{\Sigma^T}+{\Sigma^L}\right]G_0
\label{Stat2}
\end{align}
and hold for any gauge. The first of Eqs.(\ref{Stat2}) shows that the optimal propagator $D\>_{ab}^{\mu\nu}$ must be
diagonal in color indices.

In Landau gauge ($\xi\to 0$), according to Eq.(\ref{DM}) the massive propagator ${D_M}\>^{\mu\nu}$ becomes 
transversal so that any
longitudinal term in the polarization does not play any role in the first of Eqs.(\ref{Stat2}),
yielding a pure transversal solution for the optimal propagator $D\>_{ab}^{\mu\nu}$.
Moreover in the second of Eqs. (\ref{Stat2}) the longitudinal term $\Sigma^L$ is zero according to
Eq.(\ref{SigmaTL2}) and we can drop it, yielding a decoupled set of integral equations for the transversal
components. With the same notation of Eq.(\ref{Pproj}) the transversal component of the 
proper polarization $\Pi^\star_2$ in Eq.(\ref{Pistar})
can be written as
\BE
{\Pi^\star_2}^T\>_{ab}^{\mu\nu} (p)=\delta_{ab} t^{\mu\nu}(p){\Pi^\star_2}^T (p)
\label{PiL}
\EE
where the scalar function ${\Pi^\star_2}^T (p)$ is the sum of all the transversal second-order irreducible polarization
graphs (2a), (2b), (2c), (2d) and (2e).
The optimal propagator can be written as
\BE
D\>_{ab}^{\mu\nu} (p)=\delta_{ab} t^{\mu\nu}(p)T(p)
\label{DL}
\EE
where the scalar function $T(p)$ satisfies with $G(p)$ the coupled set of stationary equations (\ref{Stat2})
that in Landau gauge become
\begin{align}
T(p_E)&=\frac{1}{p_E^2+M^2}\left[1-\frac{{\Pi^\star_2}^T (p_E)}{p_E^2+M^2}\right]\nn\\
G(p_E)&=\frac{1}{-p_E^2}\left[1-\frac{\Sigma^T (p_E)}{-p_E^2}\right]
\label{StatL}
\end{align}
and we have made use of the explicit expressions of $G_0$ and $D_M$ in the Euclidean formalism.
The last equation can  be written in terms of the dressing function Eq.(\ref{ghdress}) as simply
as
\BE
\chi (p_E)=1+\frac{\Sigma^T (p_E)}{p_E^2}.
\EE

The gauge-dependent mass parameter $M$ is given by the one-loop graph (1b) and follows from its definition
in Eq.(\ref{gap1}) which closes the set of equations and must be evaluated by insertion of the
actual propagator $T(p)$ instead of the first-order massive propagator that was used in Eq.(\ref{gap2}) for the GEP.
Of course, the mass parameter does depend on the choice of gauge, as it was obvious at first order. Thus, it defines
an energy scale that must be not confused with the actual gluon mass of the renormalized propagator. 
Some typical values of the mass parameter are reported in Table I and compared with the corresponding values
in Feynman gauge as discussed below.
In Landau gauge the gap equation (\ref{gap1}) reads
\BE
M^2=\frac{N g^2 (d-1)^2}{d} I^T_{1,0}
\label{gapL}
\EE
where the integral $I^T_{1,0}$ is defined in Eq.(\ref{Inm}) and is a functional of the unknown full propagator $T(p)$. 
The stationary equations (\ref{StatL}) together with the gap equation (\ref{gapL}) can be made finite
by a proper regularization scheme and solved numerically. Details on the numerical calculation  for $d=4$
are reported in the Appendix.

\begin{table}[ht]

\centering 
\begin{tabular}{c c c } 
\hline\hline 
$g$ & \qquad $M_L$ (MeV)\quad & \qquad $M_F$ (MeV) \quad \\ [0.5ex] 
\hline 
0.5 & 172 & 457 \\ 
1.0 & 187 & 573 \\ [1ex] 
\hline 
\end{tabular}
\label{table} 
\caption{Mass parameter in Landau gauge ($M_L$) and in Feynman gauge ($M_F$) for $d=4$ and $N=3$.} 
\end{table}

\begin{figure}[t] \label{fig:TRlog}
\centering
\includegraphics[width=0.35\textwidth,angle=-90]{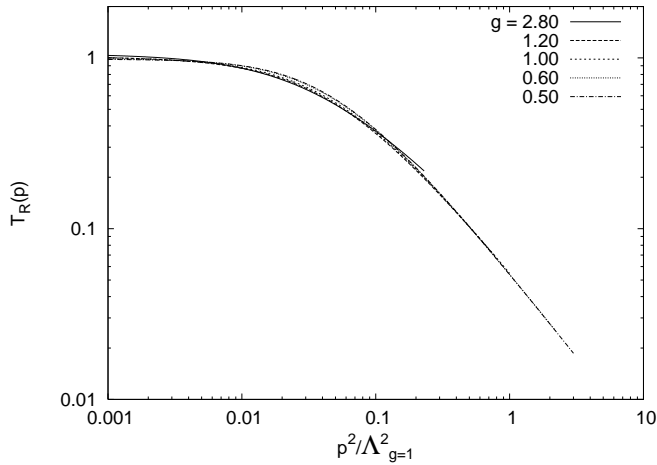}
\caption{Log-log plot of the renormalized propagator $T_R(p)$ as obtained by appropriate scaling of the bare propagator
for $N=3$, $d=4$ and for a bare coupling $g=0.50, 0.60, 1.00, 1.20, 2.80$.  
The scale is arbitrary because of scaling:
all curves have been scaled in order to fall on top of the $g=1$ bare propagator.
Energy is in units of $\Lambda_{g=1}$ so that $g(1)=1$ (for $g=1$ the curve is not rescaled).
Some deviations from scaling become more evident at the very large coupling $g=2.8$.}
\end{figure}

\begin{figure}[t] \label{fig:T05}
\centering
\includegraphics[width=0.35\textwidth,angle=-90]{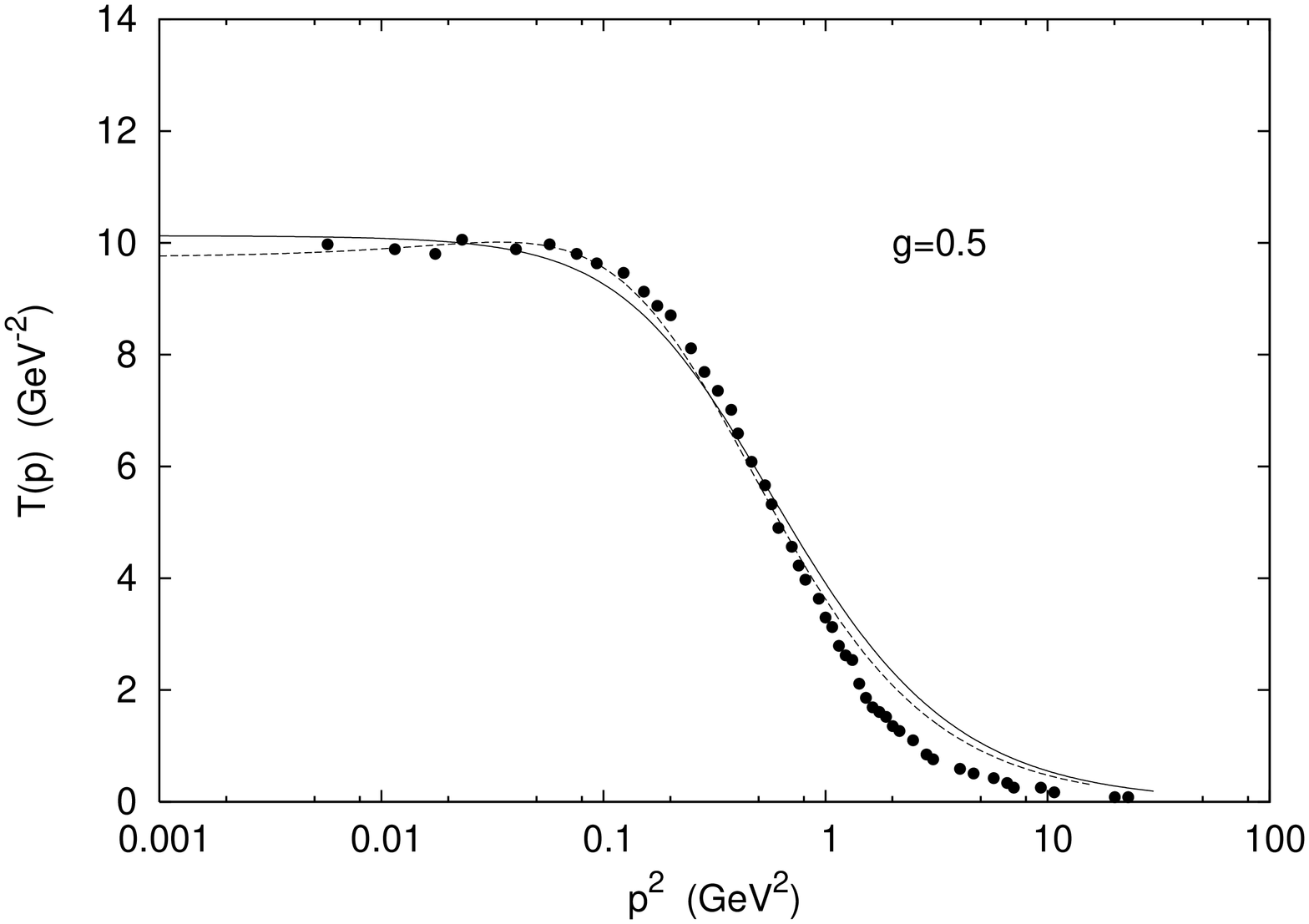}
\caption{The renormalized propagator $T{(p)}$ in Landau gauge for $N=3$, $d=4$ and for a bare coupling 
$g=0.5$. The scale has been fixed in order to fit the lattice data of Ref.\cite{bogolubsky} ($g=1.02$, L=96)
that are displayed as filled circles. The propagator in Feynman gauge is shown for comparison as a dotted line.}
\end{figure}

\begin{figure}[t] \label{fig:T10}
\centering
\includegraphics[width=0.35\textwidth,angle=-90]{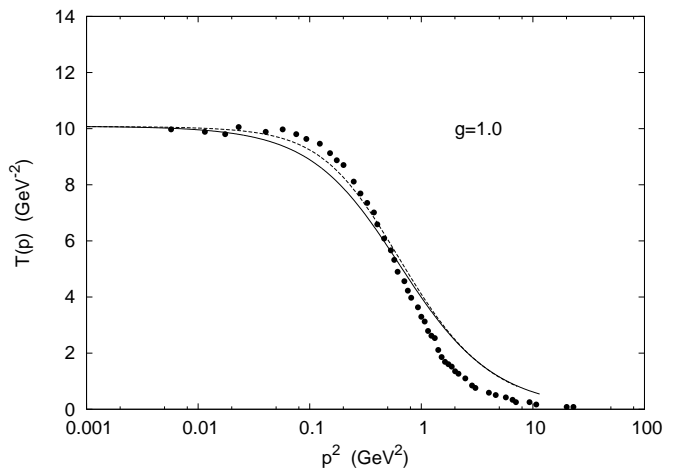}
\caption{The renormalized propagator $T{(p)}$ in Landau gauge for $N=3$, $d=4$ and for a bare coupling 
$g=1.0$. The scale has been fixed in order to fit the lattice data of Ref.\cite{bogolubsky} ($g=1.02$, L=96)
that are displayed as filled circles. The propagator in Feynman gauge is shown for comparison as a dotted line.}
\end{figure}

In this work the integrals are regularized by a finite cutoff $\Lambda$ in the Euclidean four-dimensional space 
($d=4$) where we take $p_E^2<\Lambda^2$. 
The simple choice of a cutoff gives
physical results that are directly comparable with the outcome of lattice simulations where 
a natural cutoff is provided by the lattice spacing.
The bare coupling  $g=g(\Lambda)$ is supposed to 
depend on the energy scale $\Lambda$ and
RG invariance requires that
the physical content of the theory should be invariant for a change of scale $\Lambda\to \Lambda^\prime$ followed by
a change of coupling $g\to g^\prime$. Then,
physical renormalized functions that do not depend on the cutoff can be obtained by scaling.
It is important to point out that the present regularization scheme does not need the inclusion of any
counterterm in the Lagrangian and  especially mass counterterms that are forbidden by the gauge invariance of
the Lagrangian. The interaction strength $g$ at a given scale $\Lambda$ is
the only free parameter while the function $g(\Lambda)$ can be determined by RG invariance. 
In principle, one could fix the scale by a comparison with experimental data. However, in the present
model calculation on pure Yang-Mills theory, we will fix the scale by a comparison with the lattice data.
Since the original Lagrangian does not contain any scale, it is useful to take $\Lambda=1$ and work in
units of the cutoff, at a given bare interaction strength $g$. Thus the choice of $\Lambda$ will be
equivalent to a choice of the energy units. 
RG invariance requires that a renormalized propagator $T_R (p)$ can be defined at an arbitrary scale
$\mu$ by {\it multiplicative} renormalization, that is equivalent to say that by scaling all bare functions
at different couplings can be put one on top of the other.
Of course, since the approximations and the numerical integration spoil the scaling properties, we will
consider the scaling as a test for the accuracy of the whole procedure.
We can study the scaling behaviour in a log-log plot where the bare functions should go one on top of the
other by a simple translation of the axes.

In Fig.3 the renormalized gluon propagator is shown for $N=3$, $d=4$ and for several couplings in the range $g=0.5-2.8$.
Scaling is rather good in the range $g=0.5-1.2$, but gets spoiled at the rather large coupling $g=2.8$.
That could be a limit of the second-order approximation. 

For any coupling, the energy scale can be fixed by comparison with the lattice data,  
yielding a physical renormalized propagator $T(p)$ that is shown in Fig.4 and Fig.5 for
$g=0.5$ and $g=1.0$ respectively.

As shown in Table I, when expressed in physical units the mass parameter $M$ is almost constant 
with respect to changes of the coupling, while
it remains very sensitive to the choice of gauge. However, after scaling, we can define a {\it physical} mass
$m^2=T(0)^{-1}$ that does not depend on scale and gauge because 
it is made to coincide with the lattice value $m\approx 320$ MeV.

\begin{figure}[t] \label{fig:CFR10}
\centering
\includegraphics[width=0.35\textwidth,angle=-90]{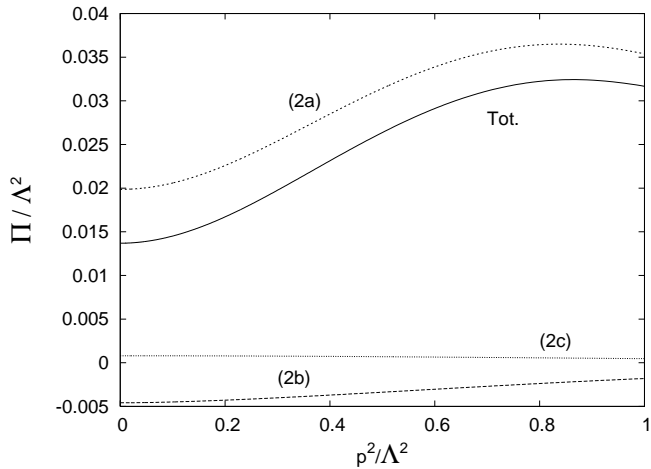}
\caption{The total polarization is displayed together with the second-order
contributions of the one-loop graphs (2a), (2b) and of the two-loop graph (2c) for $N=3$, $d=4$. 
The coupling is $g=1$ ($\beta=6$). The total polarization
includes the constant contributions $\Pi_{(2d)}=-5.25\cdot 10^{-3}$ and $\Pi_{(2e)}=2.87\cdot 10^{-3}$
in units of the cutoff.}
\end{figure}

\begin{figure}[t] \label{fig:CFR28}
\centering
\includegraphics[width=0.35\textwidth,angle=-90]{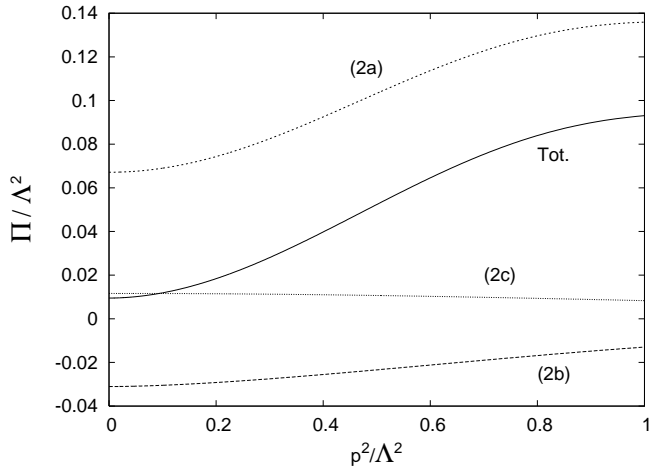}
\caption{Same as Fig.6 but for a strong coupling $g=2.8$. The constant terms are
$\Pi_{(2d)}=-7.30\cdot 10^{-2}$ and $\Pi_{(2e)}=3.48\cdot 10^{-2}$
in units of the cutoff.}
\end{figure}

A direct comparison of results in Feynman gauge\cite{varqcd} and Landau gauge shows that the renormalized 
propagator is not very sensitive to the choice of gauge, confirming that gauge invariance
survives albeit approximately. Thus the effects of a different mass scale, more than double in Feynman gauge, are
absorbed by renormalization. As shown in Fig.4 and Fig.5, where the propagator in Feynman gauge is reported
for comparison, we can say that the difference between the lattice data and the calculated curves cannot be
ascribed to gauge differences but is probably a consequence of the finite order (second order) of the approximation
or of the finite size of the lattice.
In fact, even if optimized by a variational method, the nature of the calculation is perturbative and can be
improved by inclusion of higher orders.
Actually, we do not expect that a perfect agreement could be reached in the UV limit of standard perturbation theory 
because of the simple renormalization 
scheme of the present calculation that is based on an energy cutoff. In that scheme, the spurious quadratic divergence
would spoil the result for the UV limit.
The problem of cancelling that divergence without affecting the IR limit has been discussed by several authors and
recently reviewed in Ref.\cite{huber15L}. It is a major problem that has not found a satisfactory solution yet.

Finally, it is instructive to compare the weight of the single graphs in the second order polarization 
for the studied case $N=3$, $d=4$. 
At a  rather strong coupling $g=1$ ($\beta=6$), the total polarization is reported in Fig.6 together with the
contributions of the ono-loop graphs (2a), (2b) and of the two-loop graph (2c). The total polarization
includes the constant contributions, $\Pi_{(2d)}=-5.25\cdot 10^{-3}$ and $\Pi_{(2e)}=2.87\cdot 10^{-3}$
in units of the cutoff, that are not negligible compared to the one-loop graphs.
We observe that the
two-loop term (2c) is very small and rather constant so that it could be neglected without serious consequences.
On the other hand, at the very strong coupling $g=2.8$, Fig.7 shows that the two-loop graph is still rather constant
but becomes quite important in the low energy limit where it is as large as the total polarization.
The constant terms are also rather large and amount to $\Pi_{(2d)}=-7.30\cdot 10^{-2}$ and $\Pi_{(2e)}=3.48\cdot 10^{-2}$
in units of the cutoff.

\section{Discussion}

By the explicit knowledge of the second-order polarization, several variational
strategies can be set up for the optimization of the perturbation expansion.
The method of stationary variance\cite{sigma,sigma2,gep2,varqed,varqcd} has been discussed
in the previous section. The generalized perturbation theory can also be optimized by
other methods like minimal sensitivity\cite{minimal} or by the self-consistent requirement
of a vanishing self energy. Here, we give a brief description and comparison of some different methods.

As discussed in a recent paper\cite{gep2}, 
the stationary conditions for Stevenson's method of minimal sensitivity\cite{minimal} can be written as
simply as
\BE
\Pi_2=0, \qquad \Sigma_2=0
\label{minimal}
\EE
to be compared with Eqs.(\ref{variance}) for the stationary variance.
Eqs.(\ref{minimal}) are equivalent to the requirement that the second-order effective potential
is stationary for any variation of the trial propagators.
We explored this method in Landau gauge, but Eqs.(\ref{minimal}) have no physical
solution for $N=3$ and $d=4$. In fact it is well known that sometimes that method does not show any range of
parameters where the effective potential is stationary. A second derivative would be required for imposing
that the sensitivity is minimal.

\begin{figure}[t] \label{fig:T2}
\centering
\includegraphics[width=0.35\textwidth,angle=-90]{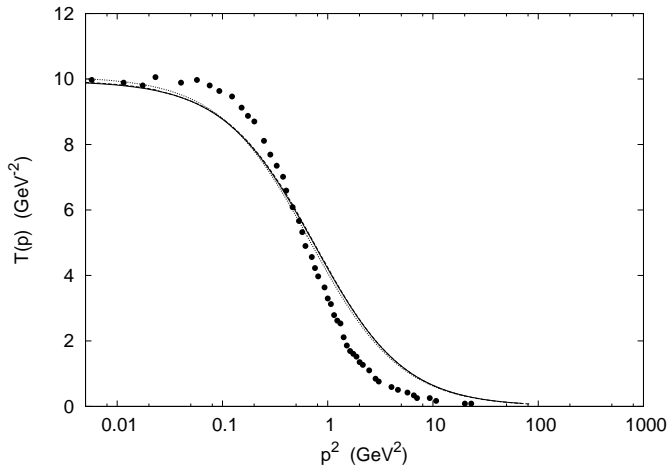}
\caption{The self-consistent renormalized propagator $T{(p)}$ in Landau gauge for $N=3$, $d=4$. 
The scale has been fixed in order to give a rough fit of the lattice data of Ref.\cite{bogolubsky} ($g=1.02$, L=96)
that are displayed as filled circles. The second-order propagator of Ref.\cite{varqcd} in Feynman 
gauge is shown for comparison as a dotted line.}
\end{figure}

An other simple approach would be based on a first-order optimization  of the expansion,
followed by a second-order evaluation of polarization and propagator\cite{tedesco}.
That would be equivalent to taking the trial propagator equal to the massive propagator 
$D_M=(-p^2+M^2)^{-1}$.
Even if the mass would not be dynamical in the trial propagator, a second order propagator
$D_2$ can be defined as usual by Dyson equations
\BE
D_2^{-1}(p)=D^{-1}(p)-\Pi_1(p)-\Pi_2^\star (p)
\label{D2}
\EE
where $D(p)$ is the trial propagator and $D=D_M$ in the actual case. 
Since the first-order optimization is self-consistent, it requires that $\Pi_1=0$ and 
in Landau gauge by Eqs.(\ref{PiL}),(\ref{DL}) we can define a transversal second-order propagator
$T_2$ that reads
\BE
T_2(p)=\left[-p^2+M^2-{\Pi_2^\star}^T (p)\right]^{-1}
\EE
Thus the second order propagator would acquire a dynamical mass that in the low energy limit
tends to $m^2=M^2-{\Pi_2^\star}^T (0)$.
The advantage of this basic approach is that the trial propagator is simple and the theory can be
renormalized by standard dimensional regularization. A similar massive model has been recently
studied\cite{tissier10,tissier11,tissier14} and shown to be in close agreement with the lattice data. 
However the mass was regarded as a free parameter rather than a variational parameter.
It would be interesting to see how close the result would be for the variational approach that
does not contain free parameters at all.

The simple first-order optimization would not be self-consistent at second order but one can
attempt and extend it by a self-consistent approach.
Eq.(\ref{D2}) is quite general and can be made self-consistent by the simple requirement that
the total proper polarization vanishes exactly
\BE
\Pi_1(p)+\Pi_2^\star (p)=0.
\label{self}
\EE
That would generalize the first-order stationary condition $\Pi_1=0$ which holds for the GEP .
If the polarization were not truncated at the second order, Eq.(\ref{self}) would be the exact
condition that the trial propagator must satisfy in order to be the exact one.
The method would be equivalent to Dyson-Schwinger equations.
Of course, truncation spoils it and the approximation depends on the accuracy of the polarization
function that can be evaluated up to second order in any gauge by the explicit expressions of the
present paper.

In Landau gauge, Eq.(\ref{self}) can be solved numerically as an integral equation for the trial propagator
$T(p)$, yelding a self-consistent second-order propagator that satisfies $T_2(p)=T(p)$
because of Eq.(\ref{D2}). The propagator satisfies a perfect scaling and can be fitted with good
accuracy by the simple expression
\BE
T(p_E)\approx \frac{Z}{p_E^2+m^2}
\EE
where $Z$ and $m$ are real parameters.
In fact, that explains the perfect scaling as any form like that, with just two free parameters, 
can be scaled on top of each other by a change of units. The result is very close to that found by
Feynman gauge in Ref.\cite{varqcd}. As shown in Fig.8, after renormalization the
self-consistent propagator can be put on the lattice data, but the agreement is worse than
found by the method of stationary variance in Fig. 4 and 5 of the previous section. 
Again, the result seems to be almost gauge invariant.

In summary, while the method of stationary variance seems to be more reliable than other variational
approaches, other attempts can be devised by the knowledge of the explicit expressions for the polarization
up to second order. Once optimized, the perturbation theory does not show divergences in the infrared, while
the ultraviolet ones can be cured by  standard regularization techniques. The explicit expressions of the
present paper hold for any regularization scheme and any choice of the gauge parameter. That would suggest
a further way to optimize the expansion, with the gauge parameter and the renormalization scheme that can be 
regarded as variational parameters\cite{stevenson2012, stevenson2013}.
A comparison between the present Landau gauge calculation and previous results in Feynman gauge\cite{varqcd} 
shows that the gluon propagator is not too much sensitive to gauge changes and that the optimized expansion 
seems to be approximately gauge invariant after renormalization, which is a desirable property from the physical 
point of view, in qualitative agreement with 
some recent results by Dyson-Schwinger equations\cite{huber15g,papavassiliou15}.

\acknowledgments
The author is in debt to Joannis Papavassiliou and Hugo Reinhardt for invaluable discussions and 
their generous hospitality.

\appendix

\section{Details on the numerical integration}

For $d=4$, all numerical integrations have been calculated as successive one-dimensional integrations by
the standard Simpson method in the Euclidean space and with an energy cutoff $p_E^2<\Lambda^2$.  
Four-dimensional integrals of simple functions of $k^2_E$
are reduced to simple one-dimensional integrals before numerical integration, according to
\BE
\int_\Lambda \kkkE A(k^2_E)=\frac{1}{8\pi^2}\int_0^\Lambda A(k^2) k^3 {\rm d}k.
\EE
Four-dimensional integrals of functions of the two variables $(k_E\cdot p_E)$ and $k^2_E$ are reduced 
to two-dimensional integrals according to

\begin{align}
\int_\Lambda \kkkE &A[(k_E \cdot p_E), k^2_E]=\nn\\
&=\int_0^\Lambda\frac{y^2 {\rm d}y}{4\pi^3}
\int_{-\sqrt{\Lambda^2-y^2}} ^{\sqrt{\Lambda^2-y^2}} A[(xp_E), (x^2+y^2) ] {\rm d}x.
\label{2dim}
\end{align}

The eight-dimensional integral of the two-loop sunset graph $(2c)$ Eq.(\ref{Pi2cA}) can be written
as a four-dimensional integral by exact integration of some variables. We notice that each single term
contributing in Eq.(\ref{Pi2cA}) can be written as
\begin{widetext}
\BE 
\int_\Lambda \kkkE\int_\Lambda \qqqE f_i\left(p_E^2,k_E^2,q_E^2,q_E\cdot(k_E+p_E), p_E\cdot k_E\right)
g_i\left(p_E^2,k_E^2,q_E^2,q_E\cdot k_E,q_E\cdot p_E, p_E\cdot k_E\right)
\label{I1}
\EE
where the function $f_i$ has one argument less than the function $g_i$, and $p$ is the external momentum. 
Let us introduce the vector $V=k_E+p_E$ and split the four-vector $q$ as the sum of two orthogonal two-dimensional
vectors $(q_1, q_2)$ and $(q_x, q_y)$ that are orthogonal and parallel to the $k-p$ plane, respectively.
Moreover we take the direction $q_y$ to be parallel to the direction of $V$. Omitting the variables
$p_E^2$,$k_E^2$,$V^2$ that are constant in the internal integration, the integral in Eq.(\ref{I1}) reads
\BE 
\int_\Lambda \kkkE \frac{1}{2(2\pi)^4}
\int_0^{\Lambda^2} {\rm d} q^2 
\int_0^{2\pi} {\rm d}\phi
\int_{-q}^{q} {\rm d} q_y
f_i(q^2, q_y)
\int_{-\sqrt{q^2-q_y^2}}^{\sqrt{q^2-q_y^2}} {\rm d} q_x
g_i(q^2,q_x,q_y)
\label{I2}
\EE
where $\tan\phi=q_2/q_1$ and $q^2=q_E^2$. The angle $\phi$ can be integrated exactly yielding a factor of $2\pi$
and denoting by $\tilde g_i(q^2,q_y)$ the integrated function
\BE
\tilde g_i(q^2,q_y)=\int_{-\sqrt{q^2-q_y^2}}^{\sqrt{q^2-q_y^2}} {\rm d} q_x
g_i(q^2,q_x,q_y)
\EE
according to Eq.(\ref{Landau2c}),
in Landau gauge we can write the transverse polarization term $\Pi_{(2c)}^T$ as
\BE
\Pi_{(2c)}^T(p)=\frac{(Ng^2)^2}{32\pi^3}\int_\Lambda \kkkE
\int_0^{\Lambda^2} {\rm d} q^2 
\int_{-q}^{q} {\rm d} q_y h(q^2,q_y)
\sum_{i=1}^{11} f_i(q^2, q_y) \tilde g_i(q^2,q_y)
\EE
where the function $h$ is
\BE
h(q^2,q_y)=T(q_E)T(k_E)T(k_E+q_E+p_E)
\EE
that only depends on $q_E\cdot V\sim q_y$ and $q^2=q_E^2$ in the internal integration.
The eleven functions $g_i$ turn out to be polynomials, and the integrated functions $\tilde g_i$ can be
evaluated exactly so that we are left with a two-dimensional internal integration and the result can only depend
on $k^2_E$ and $k_E\cdot p$. Then, the external integration follows by Eq.(\ref{2dim}), yielding
a total four-dimensional integration to be evaluated numerically.
The eleven terms $f_i$, $g_i$, $\tilde g_i$ follow from inspection of Eq.(\ref{Landau2c}). Dropping the $E$ in
the Euclidean vectors, denoting by  $s=\sqrt{q^2-q_y^2}$ and by $k_x$,$k_y$,$p_x$,$p_y$ the components of $k_E$, $p_E$ 
that are  parallel to $q_x$ and $q_y$, respectively, we can write
\begin{align}
f_1&=4, \qquad\qquad \tilde g_1=2s\nn\\
f_2&=\frac{1}{q^2k^2}, \qquad\qquad \tilde g_2=I_2(k,k)\nn\\
f_3&=\frac{1}{q^2p^2},  \qquad\qquad \tilde g_3=I_2(p,p)\nn\\
f_4&=\frac{q^2+Vq_y}{k^2q^2(k+q+p)^2},  \qquad \tilde g_4=I_2(k,k)+(V\cdot k)I_1(k)\nn\\
f_5&=\frac{-(p\cdot k)}{k^2q^2p^2},  \qquad\qquad \tilde g_5=I_2(p,k)\nn\\
f_6&=\frac{2}{p^2(k+p+q)^2},  \qquad \tilde g_6=I_2(p,p)\nn\\
f_7&=\frac{4(p\cdot V)}{p^2(k+p+q)^2},  \qquad \tilde g_7=I_1(p)\nn\\
f_8&=\frac{2(p\cdot V)^2}{p^2(k+p+q)^2},  \qquad \tilde g_8=2s\nn\\
f_9&=\frac{-2(p\cdot k)}{p^2k^2(k+p+q)^2},  \qquad \tilde g_9=I_2(k,p)+(k\cdot V)I_1(p)
+(p\cdot V) I_1(k)+(2s)(p\cdot V)(k\cdot V)\nn\\
f_{10}&=\frac{1}{p^2k^2q^2(k+p+q)^2},  \qquad \tilde g_{10}=(p\cdot V)^2I_2(k,k)+I_4(k,p)+2(p\cdot V)I_3(k,k,p)\nn\\
f_{11}&=\frac{-(k\cdot p)(q^2+Vq_y)}{p^2k^2q^2(k+p+q)^2},  \qquad \tilde g_{11}=I_2(k,p)+(p\cdot V)I_1(k)
\label{fg}
\end{align}
where the functions $I_n$ are defined as
\begin{align}
I_1(k)&=(2s)q_yk_y\nn\\
I_2(k,p)&=(2s)\left[q_y^2k_yp_y+\frac{1}{3}s^2k_xp_x\right]\nn\\
I_3(k,k,p)&=\frac{2}{3}s^3q_y(k_x^2p_y+2k_x k_yp_x)+(2s)q_y^3(k_y^2p_y)\nn\\
I_4(k,p)&=\frac{2}{5}s^5(k_x^2p_x^2)+\frac{2}{3}s^3 q_y^2(k_x^2p_y^2+k_y^2p_x^2+4k_xk_yp_xp_y)
+(2s)q_y^4(k_y^2p_y^2)
\label{In}
\end{align}
\end{widetext}

\end{document}